\documentclass[a4paper]{article}

\usepackage[dvipdfmx]{graphicx, color}
\usepackage{wrapfig}
\usepackage{amsmath, amssymb}
\usepackage{bm}
\usepackage{pifont}
\usepackage{cite}
\usepackage[dviout]{pict2e}
\usepackage{booktabs}

\begin{document}

\title{Full Simulation Study of the Higgs Branching Ratio \\ into Tau Lepton Pairs at the ILC with $\sqrt{s}=500$~GeV}
\date{March 27, 2014}
\author{Shin-ichi Kawada$^{1,\dagger}$, Keisuke Fujii$^2$, Taikan Suehara$^3$,\\ Tohru Takahashi$^1$, Tomohiko Tanabe$^4$ }
\maketitle{}

\noindent 1: Graduate School of Advanced Sciences of Matter, Hiroshima University, 1-3-1, Kagamiyama, Higashi-Hiroshima, Hiroshima, 739-8530, Japan \\
2: High Energy Accelerator Research Organization (KEK), 1-1, Oho, Tsukuba, Ibaraki, 305-0801, Japan \\
3: Graduate School of Science, Kyushu University, 6-10-1, Hakozaki, Higashi-ku, Fukuoka, 812-8581, Japan \\
4: International Center for Elementary Particle Physics (ICEPP), The University of Tokyo, 7-3-1, Hongo, Bunkyo-ku, Tokyo, 113-0033, Japan \\ \\
$\dagger$ : \verb|s-kawada@huhep.org|

\begin{abstract}
\footnote{Talk presented at the International Workshop on Future Linear Colliders (LCWS13), Tokyo, Japan, 11-15 November 2013.}
We evaluate the expected measurement accuracy of the branching ratio of the Standard Model Higgs boson decaying into tau lepton pairs $h \to \tau ^+ \tau ^-$ at the ILC with a center-of-mass energy of $\sqrt{s}=500$~GeV with a full simulation of the ILD detector.
We assume a Higgs mass of $M_h = 125$ GeV, a branching ratio of $\mathrm{BR}(h \to \tau ^+ \tau ^-) = 6.32 \ \%$, beam polarizations of $P(e^-, e^+) = (-0.8, +0.3)$, and an integrated luminosity of $\int L \ dt = 500 \ \mathrm{fb ^{-1}}$.
The Higgs-strahlung process $e^+ e^- \to Zh$ with $Z\rightarrow q\overline{q}$ and the $WW$-fusion process $e^+ e^- \to \nu \overline{\nu}h$ are expected to be the most sensitive channels at $\sqrt{s}=500$~GeV.
Using a multivariate analysis technique, we estimate the expected relative measurement accuracy of the branching ratio $\Delta (\sigma \cdot \mathrm{BR}) / (\sigma \cdot \mathrm{BR})$ to be 4.7\% and 7.4\% for the $q\overline{q}h$ and $\nu \overline{\nu} h$ final states, respectively.
The results are cross-checked using a cut-based analysis.
\end{abstract}

\section{Introduction}

After the discovery of the Higgs boson by the LHC experiments~\cite{ATLAS, CMS}, the investigation of the properties of the Higgs boson has become one of the most important themes in particle physics.
In the Standard Model, the Yukawa coupling of matter fermions with a Higgs boson is proportional to the fermion mass.
However, new physics models predict a deviation of the Yukawa coupling from the Standard Model prediction.
The size of the deviation is expected to be small if the scale of new physics is high.
Specifically, the allowed deviation can be at the few-percent level even if no additional new particles are to be found at the LHC in the next several years~\cite{deviate}.
Since the branching ratio measurement is used as an input in the extraction of the Yukawa coupling, a precise determination of the branching ratio is essential to probe new physics.

In this study, we focus on the branching ratio of the Higgs boson decays into tau lepton pairs.
The study of the tau lepton Yukawa coupling is special in the following ways.
The mass of the tau lepton is known to a very good precision 
unlike quarks, which typically suffer from the theoretical uncertainties arising from QCD.
Also, the deviation in the lepton Yukawa coupling could well differ from the quark Yukawa coupling, such as in the lepton-specific Two-Higgs Doublet Model.
Thus, the tau Yukawa coupling is an ideal probe for new physics.

In this study, we estimate the measurement accuracy $\Delta (\sigma \cdot \mathrm{BR}) / (\sigma \cdot \mathrm{BR})$ of the $h \to \tau ^+ \tau ^-$ branching ratio at the center-of-mass energy of $\sqrt{s}=500$~GeV at the ILC with the ILD full detector simulation.
The results of $\sqrt{s} = 250$~GeV are summarized in Ref.~\cite{LCNOTE} and presented at the ECFA 2013 workshop~\cite{TALKECFA}.

\section{Signal and Background Processes}

\begin{figure}[!h]
\centering
\includegraphics[width = 16.0truecm]{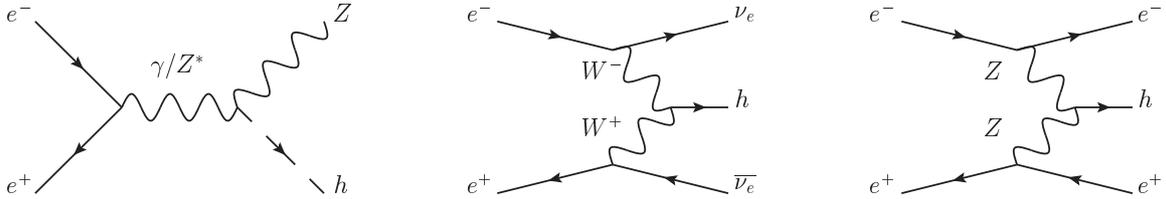}
\caption{The diagrams of the Higgs production processes.
(left): Higgs-strahlung process, (middle): $WW$-fusion process, (right): $ZZ$-fusion process.}
\label{signals}
\end{figure}

The diagrams for the Higgs production processes are shown in Figure~\ref{signals}.
At $\sqrt{s} = 500$ GeV, the $WW$-fusion process and Higgs-strahlung process are the most dominant production processes.
We analyze the $WW$-fusion process and the Higgs-strahlung process with $Z \to q\overline{q}$ decays, which are expected to be the most sensitive channels because of the high statistics.
The cross sections of the $WW$-fusion and Higgs-strahlung processes at $\sqrt{s} = 500$ GeV are 149.5~fb and 100.4~fb, respectively.

\clearpage

\begin{figure}[!h]
\centering
\includegraphics[width = 16.0truecm]{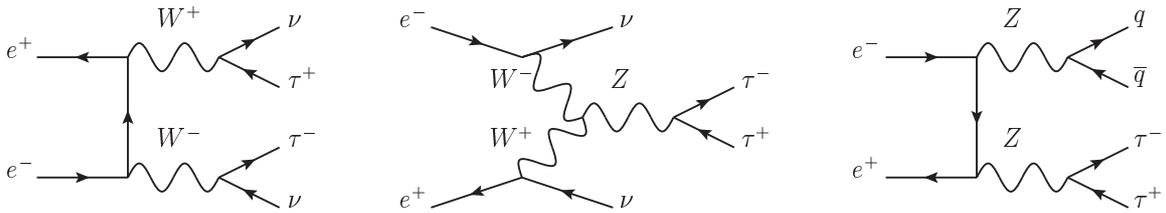}
\caption{Examples of irreducible background processes.
(left): $\nu \nu \tau ^+ \tau ^-$ background via $e^+ e^- \to W^+ W^-$ for $WW$-fusion signal, (middle): $\nu \nu \tau ^+ \tau ^-$ background via $WW$-fusion ($e^+ e^- \to \nu \nu Z$) for $WW$-fusion signal, (right): $q\overline{q}\tau ^+ \tau ^-$ background for Higgs-strahlung signal.}
\label{backgrounds}
\end{figure}

The diagrams for the main backgrounds which have the same final states as the signal
are shown in Figure~\ref{backgrounds}.
For the $WW$-fusion signal, the $\nu \nu \tau ^+ \tau ^-$ final state which proceeds via $e^+ e^- \to W^+ W^-,\,ZZ$ is an irreducible background.
In addition, the $e^+ e^- \to \nu \nu Z$ final state via the $WW$-fusion process with $Z \to \tau ^+ \tau ^-$ is another irreducible background.
For the Higgs-strahlung signal, the $e^+ e^- \to ZZ$ process is the main irreducible background to the signal.

\section{Simulation Conditions}

We assume a Higgs mass of $M_h = 125$ GeV, a branching ratio of $\mathrm{BR}(h \to \tau ^+ \tau ^-) = 6.32 \ \%$~\cite{NNLO}, an integrated luminosity of $\int L \ dt = 500 \ \mathrm{fb^{-1}}$, and beam polarizations of $P(e^-, e^+) = (-0.8, +0.3)$.

We use the signal and background samples which are prepared for the studies presented in the ILC Technical Design Report~\cite{TDR1, TDR2, TDR3, TDR4}.
The beam energy spectrum includes the effects due to beamstrahlung and the initial state radiation.
The beam-induced backgrounds from $\gamma\gamma$ interactions which give rise to hadrons are included in all signal and background processes.
The background processes from $e^+e^-$ interactions are categorized according to the number of final-state fermions: two fermions (2f), four fermions (4f), five fermions (5f), six fermions (6f).
We also include $\gamma \gamma \to $~2f and 4f processes.
The detector response is simulated using full simulation based on Geant4~\cite{Geant4} as described below, except for the $\gamma\gamma\to$~2f sample which is treated using fast detector simulation using SGV~\cite{SGV}.

We perform the detector simulation with \verb|Mokka|~\cite{Mokka}, a Geant4-based~\cite{Geant4} full simulation, with the ILD detector model \verb|ILD_o1_v05|.
\verb|TAUOLA|~\cite{TAUOLA} is used for the tau decay simulation.
The ILD detector model consists of a vertex detector, a time projection chamber, an electromagnetic calorimeter, a hadronic calorimeter, a return yoke, muon systems, and forward components.

\section{Event Reconstruction and Event Selection}
\subsection{Analysis of $\nu \overline{\nu} h$ mode at $\sqrt{s} = 500$ GeV}

First, in order to reduce the particles due to the beam-induced background, we apply the $k_T$ clustering algorithm~\cite{kT1, kT2}, as implemented in the \verb|FastJet|~\cite{FastJet} package, with a jet radius $R$ of 1.0.
The particles in the forward region which are typically due to the beam-induced background are not included in the resulting jet objects.
We make a list of particles which are used to form these jets, which are used as a pool of particles within which to look for tau lepton decays.
Our tau finder searches for the charged track with the highest energy among the remaining particles, and combines the neighboring particles within an angle of less than 0.76~rad with respect to the energetic track, provided that the combined mass is less than 2~GeV.
The resulting object consisting of one or three tracks, with additional neutral clusters, is regarded as a tau lepton candidate and is set aside.
This procedure is repeated until there are no charged particles left in the list of particles.
The most energetic $\tau ^+$ and $\tau ^-$ candidates are combined into a Higgs boson candidate.

\begin{figure}[!h]
\centering
\includegraphics[width = 10.0truecm]{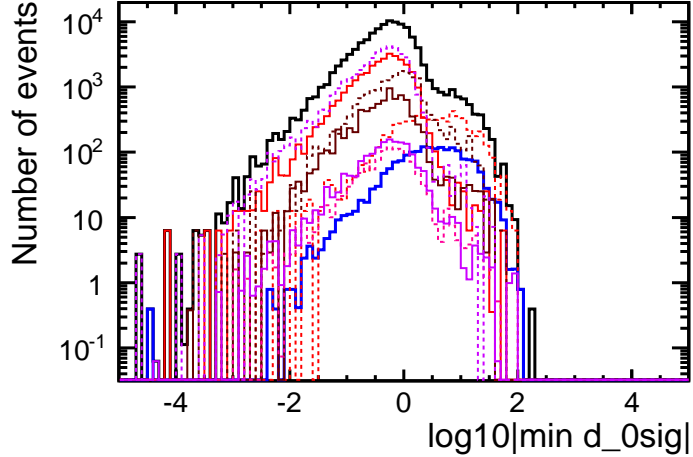}
\caption{The distributions of $\log _{10}|\min (d_0 \mathrm{sig})|$ after all but the $\log _{10}|\min (d_0 \mathrm{sig})|$ cuts.
The black histogram is the sum of the signal and all the backgrounds, while the blue one shows the signal, the solid red for $\nu \nu \ell \ell$ background, the dotted brown for $\nu \nu \tau \ell$ background, the dotted red for $\nu \nu \tau \tau$ background, the solid brown for other 4f background, the dotted purple for $\gamma \gamma \to$ 2f background, the solid purple for $\gamma \gamma \to$ 4f background processes, respectively.
}
\label{d0sig}
\end{figure}

We describe the procedures for the cut-based analysis first.
Preselections are applied to the number of tau candidates $\tau ^{+(-)} \ge 1$ and the number of tracks $\le 6$ in the event after the removal of beam-induced background.
Then we apply a second preselection to suppress huge $\gamma \gamma \to$ 2f process with the following cuts; the number of energetic tracks, where energetic means having a transverse momentum greater than 5 GeV, has to be one or more and the total visible transverse momentum $P_t$ of the final state has to satisfy $P_t > 10$ GeV.
After the preselections, we apply the following cuts sequentially to maximize the signal significance; $M_{\mathrm{vis}} < 130$ GeV, $E_{\mathrm{vis}} < 215$ GeV, $P_t > 50$ GeV, 0.7 $<$ thrust $<$ 0.97, $|\cos \theta _{\mathrm{miss}}| < 0.89$, 25 GeV $< M_{\tau ^+ \tau ^-} < 115$ GeV, $E_{\tau ^+ \tau ^-} > 35$ GeV, $-0.81 < \cos \theta _{\tau ^+ \tau ^-} < 0.55$, $\cos \theta _{\mathrm{acop}} < 0.96$, and $\log _{10} |\min (d_0 \mathrm{sig})| > 0.4$, where $M_{\mathrm{vis}}$ is the visible mass, $E_{\mathrm{vis}}$ is the visible energy, $P_t$ is the transverse momentum, $\theta _{\mathrm{miss}}$ is the angle of missing momentum with respect to the beam axis, $M_{\tau ^+ \tau ^-}$ is the invariant mass of tau pair system, $E_{\tau ^+ \tau ^-}$ is the energy of tau pair system, $\theta _{\tau ^+ \tau ^-}$ is the angle between $\tau ^+$ and $\tau ^-$, $\theta _{\mathrm{acop}}$ is the acoplanarity angle between $\tau ^+$ and $\tau ^-$, and $\min (d_0 \mathrm{sig})$ is the smaller $d_0$ impact parameter divided by the error of $d_0$ between $\tau ^+$ and $\tau ^-$, respectively.
Figure~\ref{d0sig} shows the $\log _{10} |\min (d_0 \mathrm{sig})|$ distribution for the final sample (black line) and its breakups to signal and various background contributions.

After all the cuts, the signal events of 1036 and the background events of 6872 are remained.
The statistical signal significance is calculated to be $S / \sqrt{S + B} = 1036 / \sqrt{1036 + 6872} = 11.7 \sigma$.
This result corresponds to the precision of $\Delta (\sigma \cdot \mathrm{BR}) / (\sigma \cdot \mathrm{BR}) = 8.5\%$.

\begin{figure}[!h]
\centering
\includegraphics[width = 10.0truecm]{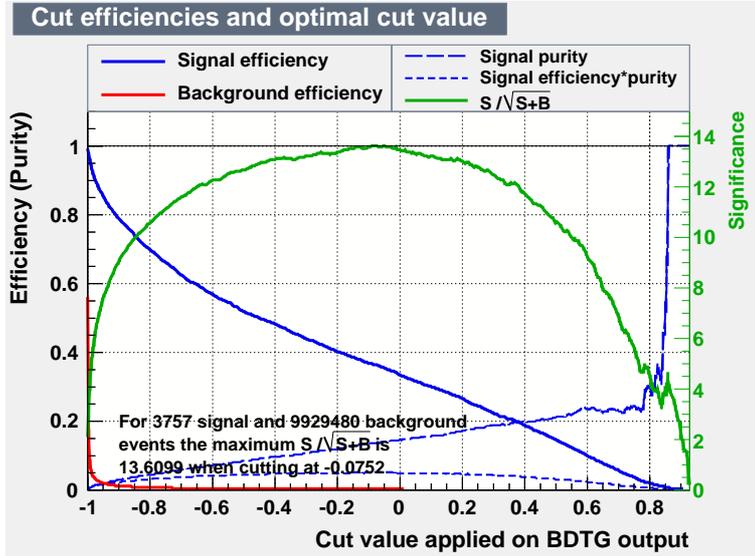}
\caption{The multivariate classifier response for $\nu \overline{\nu} h$ mode.
The green plot shows the signal significance.}
\label{BDTG_case02}
\end{figure}

We now describe the multivariate analysis.
We use the TMVA package in ROOT~\cite{ROOT}.
As a preselection,
we apply the following cuts to suppress trivial backgrounds; the number of $\tau ^{+(-)} \ge 1$, the number of tracks $\le 6$, the number of energetic tracks $\ge 1$, $M_{\mathrm{vis}} < 135$ GeV, $P_t > 10$ GeV, and thrust $> 0.7$.
We use the following 12 parameters as the inputs to the multivariate training; the number of tracks, the number of energetic tracks, $M_{\mathrm{vis}}$, $P_t$, $\max P_t$(maximum transverse momentum of track in an event), $\cos \theta _{\mathrm{miss}}$, thrust, $M_{\tau ^+ \tau ^-}$, $E_{\tau ^+ \tau ^-}$ (energy of tau pair system), $\cos \theta _{\tau ^+ \tau ^-}$, $\cos \theta _{\mathrm{acop}}$, and $\log _{10} |\min (d_0 \mathrm{sig})|$.
Figure~\ref{BDTG_case02} shows the response of the multivariate classifier.
The 1361 signals and 8648 background are left when applying the cut which extracts the maximum significance.
The signal significance is calculated to be $S / \sqrt{S +B} = 1361 / \sqrt{1361 + 8648} = 13.6\sigma$.
This means that the precision of $\Delta (\sigma \cdot \mathrm{BR}) / (\sigma \cdot \mathrm{BR}) = 7.4\%$.
The multivariate analysis improves the result by 15\% compared with the cut-based analysis.

\subsection{Analysis of $q \overline{q} h$ mode at $\sqrt{s} = 500$ GeV}

In this mode, the tau lepton candidates are reconstructed first, followed by the dijet reconstruction of the $Z$ decay.
We apply the $k_T$ algorithm~\cite{kT1, kT2} with the jet radius $R$ of 1.2 to remove beam-induced backgrounds before starting event reconstruction.

First, we apply the tau finder to the remaining objects.
This tau finder searches the highest energy track and combine the neighboring particles, which satisfy $\cos \theta _{\mathrm{cone}} > 0.98$, with the combined mass less than 2 GeV ($\theta _{\mathrm{cone}}$: the cone angle with respect to the highest energy track).
We regard the combined object as a tau candidate.
Then, we apply the selection cuts as following; $E_{\mathrm{tau \ candidate}} > 3$ GeV, $E_{\mathrm{cone}} < 0.1\times E_{\mathrm{tau \ candidate}}$ with $\cos \theta ' _{\mathrm{cone}} = 0.9$, and rejecting 3-prong events with neutral particles, where $E_{\mathrm{tau \ candidate}}$ is the energy of a tau candidate, $E_{\mathrm{cone}}$ is the cone energy of a tau candidate with the cone angle of $\theta ' _{\mathrm{cone}}$ ($\theta ' _{\mathrm{cone}}$: the cone angle with respect to a tau candidate).
These selections are tuned to minimize the misidentification of fragments of quark jets as tau decays.
After the selection, we apply the charge recovery to obtain better efficiency.
The charged particles in a tau candidate which have the energy less than 2 GeV are detached one by one, the one with the smallest energy first, until the following conditions are met; the charge of a tau candidate is $\pm 1$, and the number of track(s) in a tau jet is 1 or 3.
After the selection and detaching, we repeat the above processes until there are no charged particles which have the energy greater than 2 GeV.

After finishing the tau reconstruction, we apply the collinear approximation~\cite{colapp} to reconstruct the invariant mass of tau pair system.
In this approximation, we assume that the visible decay products of the tau lepton and the neutrino(s) from the tau decay is collinear, and the contribution of the missing transverse momentum comes only from the neutrino(s) from tau decay.

After the approximation, we apply the Durham jet clustering~\cite{Durham} with two jets for the remaining objects to reconstruct $Z$ boson.

\begin{figure}[!h]
\centering
\includegraphics[width = 10.0truecm]{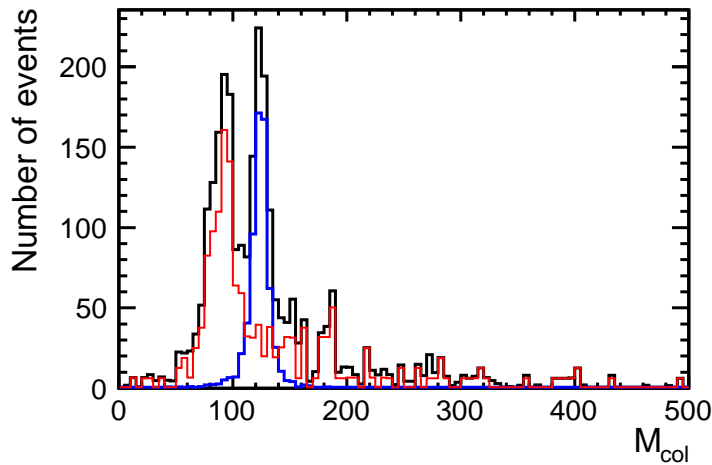}
\caption{The distribution of $M_{\mathrm{col}}$ in the sequential cuts.
Black, blue, red histograms show the summing up of all processes, signal, 4f background process, respectively.}
\label{mass_colapp_kt}
\end{figure}

We describe the procedures for the cut-based analysis first.
Before optimizing the cuts, we apply the preselection as follows; the number of quark jets $= 2$, the number of $\tau ^{+(-)} = 1$, the number of tracks $\ge 9$, $M_{\mathrm{col}} > 0$ GeV, and $E_{\mathrm{col}} > 0$ GeV, where $M_{\mathrm{col}} (E_{\mathrm{col}})$ is the invariant mass (energy) of tau pair system with collinear approximation.
We apply the following cuts to extract maximum significance; $P_t > 190$ GeV, thrust $< 0.93$, $|\cos \theta _{\mathrm{miss}}| < 0.96$, 80 GeV $< M_Z (M_{q\overline{q}}) < 145$ GeV, $E_Z (E_{q\overline{q}}) > 190$ GeV, $M_{\tau ^+ \tau ^-} < 125$ GeV, $E_{\tau ^+ \tau ^-} < 235$ GeV, $\cos \theta _{\tau ^+ \tau ^-} < 0.58$, $\log _{10}|d_0\mathrm{sig} (\tau ^+)| + \log _{10}|d_0\mathrm{sig} (\tau ^-)| > 0.2$, 110 GeV $< M_{\mathrm{col}} < 140$ GeV, $E_{\mathrm{col}} < 290$ GeV, and $M_{\mathrm{recoil}} > 50$ GeV, where $P_t$ is the transverse momentum, $\theta _{\mathrm{miss}}$ is the angle of missing momentum with respect to the beam axis, $M_Z (M_{q\overline{q}})$ is the invariant mass of quark pair system, $E_Z (E_{q\overline{q}})$ is the energy of quark pair system, $M_{\tau ^+ \tau ^-} (E_{\tau ^+ \tau ^-})$ is the invariant mass (energy) of tau pair system, $\theta _{\tau ^+ \tau ^-}$ is the angle between $\tau ^+$ and $\tau ^-$, $\theta _{\mathrm{acop}}$ is the acoplanarity angle between $\tau ^+$ and $\tau ^-$, $d_0\mathrm{sig} (\tau ^{+(-)})$ is the $d_0$ impact parameter divided by error of $d_0$ of $\tau ^{+(-)}$, $M_{\mathrm{col}} (E_{\mathrm{col}})$ is the invariant mass (energy) of tau pair system with collinear approximation, and $M_{\mathrm{recoil}}$ is the recoil mass against $Z$ boson, respectively.
Figure~\ref{mass_colapp_kt} shows the $M_{\mathrm{col}}$ distribution in the sequential cuts.

After all the cuts, the signal events of 548.5 and the background events of 170.5 are remained.
The statistical significance is calculated to be $S / \sqrt{S + B} = 548.5 / \sqrt{548.5 + 170.5} = 20.5 \sigma$.
This result corresponds to the precision of $\Delta (\sigma \cdot \mathrm{BR}) / (\sigma \cdot \mathrm{BR}) = 4.9\%$.

\begin{figure}[!h]
\centering
\includegraphics[width = 10.0truecm]{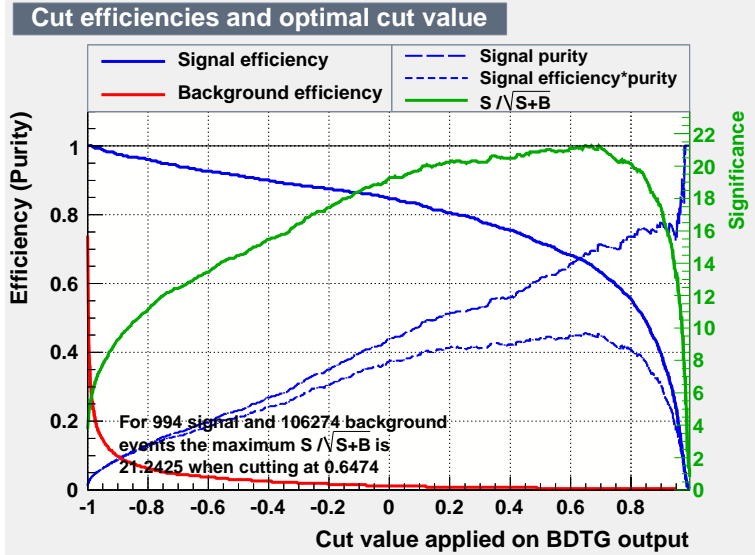}
\caption{The multivariate classifier response for $q \overline{q} h$ mode.
The green plot shows the signal significance.}
\label{BDT_case05}
\end{figure}

We now describe the multivariate analysis.
As a preselection, we apply the following cuts to reject trivial backgrounds; the number of quark jets $= 2$, the number of $\tau ^{+(-)} = 1$, 100 GeV $< M_{\mathrm{vis}} < 510$ GeV, $P_t > 80$ GeV, thrust $< 0.98$, $M_Z (M_{q\overline{q}}) > 20$ GeV, $E_Z (E_{q\overline{q}}) > 80$ GeV, $M_{\tau ^+ \tau ^-} < 140$ GeV, $M_{\mathrm{col}} > 20$ GeV, and $E_{\mathrm{col}} < 400$ GeV, where $M_{\mathrm{vis}}$ is the visible mass.
We use the following 17 parameters as the inputs; the number of tracks, $M_{\mathrm{vis}}$, $P_t$, thrust, $\cos \theta _{\mathrm{miss}}$, $M_Z (M_{q\overline{q}})$, $E_Z (E_{q\overline{q}})$, $\cos \theta _{q\overline{q}}$ ($\theta _{q\overline{q}}$: angle between quarks), $M_{\tau ^+ \tau ^-}$, $E_{\tau ^+ \tau ^-}$, $\cos \theta _{\tau ^+ \tau ^-}$, $\cos \theta _{\mathrm{acop}}$, $\log _{10}|d_0\mathrm{sig} (\tau ^+)| + \log _{10}|d_0\mathrm{sig} (\tau ^-)|$, $\log _{10}|z_0\mathrm{sig} (\tau ^+)| + \log _{10}|z_0\mathrm{sig} (\tau ^-)|$ ($z_0\mathrm{sig}$: $z_0$ impact parameter divided by the error of $z_0$), $M_{\mathrm{col}}$, $E_{\mathrm{col}}$, and $M_{\mathrm{recoil}}$.
Figure~\ref{BDT_case05} shows the response of the multivariate classifier.
The number of events surviving the event selection is 659.0 for the signal and 303.4 for the background.
The signal significance is calculated to be $S / \sqrt{S +B} = 659.0 / \sqrt{659.0 + 303.4} = 21.2\sigma$,
which implies a precision of $\Delta (\sigma \cdot \mathrm{BR}) / (\sigma \cdot \mathrm{BR}) = 4.7\%$.
The result of the multivariate analysis improved by 4\% compared with the cut-based analysis.

\section{Summary}

We evaluated the expected measurement accuracy of the branching ratio $\Delta (\sigma \cdot \mathrm{BR}) / (\sigma \cdot \mathrm{BR})$ of the $h \to \tau ^+ \tau ^-$ mode at $\sqrt{s} = 500$ at the ILC with a full simulation of the ILD detector model, assuming $M_h = 125$ GeV, $\mathrm{BR}(h \to \tau ^+ \tau ^-) = 6.32 \ \%$, $\int L \ dt = 500 \ \mathrm{fb^{-1}}$, and beam polarizations $P(e^-, e^+) = (-0.8, +0.3)$.
We analyzed the $\nu \overline{\nu} h$ and $q\overline{q}h$ final states using a cut-based approach and a multivariate approach.
The results are summarized in Table~\ref{500GeV_summary}.
\begin{table}[!h]
\centering
\caption{The analysis results of $\sqrt{s} = 500$ GeV.}
\begin{tabular}{ccc} \hline
Mode & Cut-based & Multivariate \\ \hline
$\nu \overline{\nu} h$ & 8.5\% & 7.4\% \\ 
$q \overline{q} h$ & 4.9\% & 4.7\% \\ \hline
\label{500GeV_summary}
\end{tabular}
\end{table}
The obtained performance in this study is only indicative and is expected to improve further, for example,
with more optimization in the multivariate approach.
Combining the results with the analysis of the Higgs-strahlung $Zh$ process with $Z\to \ell^+\ell^-$ decays
is expected to slightly improve the overall precision.


\end{document}